\documentclass[aps,prb,twocolumn,superscriptaddress,nofootinbib]{revtex4-1}
\usepackage{amssymb,graphicx,color}
\usepackage[intlimits]{amsmath}
\usepackage[english]{babel}
\usepackage[colorlinks]{hyperref}
\hypersetup{
colorlinks=magenta,
linkcolor=magenta,
filecolor=magenta,
urlcolor=blue,
citecolor=magenta
}
\usepackage[normalem]{ulem}
\usepackage{comment}
\usepackage{ragged2e}
\usepackage{enumerate}

\begin{document}

\title{Noise-induced dynamical localization and delocalization}
\author{Vatsana Tiwari}
\affiliation{Department of Physics, Indian Institute of Science Education and Research, Bhopal, India}
\author{Devendra Singh Bhakuni}
\affiliation{Department of Physics, Indian Institute of Science Education and Research, Bhopal, India}
\affiliation{Department of Physics, Ben-Gurion University of the Negev, Beer-Sheva 84105, Israel}
\author{Auditya Sharma}
\email{auditya@iiserb.ac.in}
\affiliation{Department of Physics, Indian Institute of Science Education and Research, Bhopal, India}

\begin{abstract}
We investigate the effect of a two-level jump process or random
telegraph noise on a square wave driven tight-binding lattice. In the
absence of the noise, the system is known to exhibit dynamical
localization for specific ratios of the amplitude and the frequency of
the drive. We obtain an exact expression for the probability
propagator to study the stability of dynamical localization against
telegraph noise. Our analysis shows that in the presence of noise, a
proper tuning of the noise parameters destroys dynamical localization
of the clean limit in one case, while it induces dynamical
localization in an otherwise delocalized phase of the clean
model. Numerical results help verify the analytical findings. A study
of the dynamics of entanglement entropy from an initially half-filled
state offers complementary perspective.
\end{abstract}

\maketitle

\section{Introduction}
A charged particle subjected to a static electric field together with
a periodic lattice potential performs bounded and oscillatory motion,
which is termed as \textit{Bloch oscillations} ~\cite{zener1934theory,
  PhysRev.117.432,PhysRevB.33.5494,dominguez2010beyond}. Here the usual Bloch band
structure of a crystal lattice is destroyed and instead, we have an
equispaced energy spectrum termed as the \textit{Wannier-Stark
  ladder}\cite{Hartmann_2004, gong2005dynamics}. Moreover, all the
single-particle wavefunctions are localized and therefore the name
`Wannier-Stark (WS) localization' is associated with this
phenomenon. Bloch oscillations have been realized in a wide variety of
physical systems such as trapped cold atoms~\cite{PhysRevLett.76.4508,
  PhysRevLett.76.4504}, semiconductor superlattices
~\cite{PhysRevLett.79.301,PhysRevLett.70.3319}, photonic systems
\cite{PhysRevLett.91.263902, PhysRevLett.83.4752}, and spin systems
using a superconducting quantum processor~\cite{Guo2021}. Furthermore,
in the presence of nearest-neighbor interactions, the system exhibits
many-body (Stark)
localization~\cite{PhysRevLett.122.040606,PhysRevB.102.054206,
  PMID:32131055}. The existence of Stark MBL has been probed
experimentally and theoretically in recent works~\cite{guo2020stark,
  morong2021observation, PhysRevB.102.054206}.  In the presence of a
time-dependent electric field, the periodic drive at specific ratios
of drive amplitude and frequency $\left(A/\omega\right)$ effectively
suppresses the hopping strength and leads to dynamical localization
\cite{ gong2005dynamics,LENZ200387,PhysRevB.34.3625,
  PhysRevLett.67.516, van2019bloch,SciPostPhys.3.4.029}. Moreover, the
combined action of ac and dc electric fields gives rise to fascinating
phenomena like coherent destruction of WS
localization~\cite{Holthaus_1995,PhysRevLett.75.3914,PhysRevB.98.045408}
and super Bloch oscillations~\cite{PhysRevB.86.075143,
  CAETANO20112770,PhysRevA.83.053627} in the non-interacting limit,
and coherent destruction of Stark MBL in the interacting
limit~\cite{PhysRevB.102.024201}. Furthermore, a coupling to bosonic heat bath (with Ohmic dissipation) leads to decoherence of the Bloch oscillations and gives rise to a dissipative transport~\cite{PhysRevB.104.125401}.

In a realistic situation, the system is almost always coupled to a
thermalizing bath; moreover the unwanted temporal fluctuations in the
drive can lead to aperiodicity, which eventually leads to dephasing
and may destroy some of the above carefully-tuned properties. Such
noise either originates from the lattice vibrations where the phonons
are randomly excited or arises due to the fluctuations in the
battery~\cite{PhysRevB.96.195435,PhysRevB.82.245417,dattagupta2012relaxation}. When
the electric field is static, this noise is associated with many
interesting features such as incoherent destruction of WS localization
and renormalization of the Bloch frequency~\cite{PhysRevB.99.155149,
  PhysRevB.101.184308} in the non-interacting limit, and the dephasing
of Stark-MBL in the interacting case~\cite{PhysRevLett.123.030602}.
In this work, we analyze a model in which a periodic electric field is
subjected to a two-level jump process or \emph{telegraph noise}.  In
particular, we investigate the effect of temporal noise on dynamical
localization.

The main findings of our work are as follows. For a system subjected
to a (telegraph) noisy time-periodic (square wave) electric field, we
obtain an exact expression for the probability propagator
$\tilde{\mathcal{P}}_{n}\left(t\right)$. In the clean limit, the well known
case of Bloch oscillations and dynamical localization are verified for
a static and periodic field respectively. Moreover, we generalize the
results for a combined ac and dc field and provide an exact expression
for the probability propagator leading to the cases of
coherent-destruction of WS localization and super-Bloch oscillations.
The rapid relaxation limit of the stochastic field is a
particular focus of our work.  Denoting the bias in the probabilities
of the two levels of the field to be $\delta p$, we show that in the
zero bias case $\left(\delta p=0\right)$, for small values of noise,
dynamical localization survives. However in the long time limit and
for large values of noise, we observe that noise decoheres the system
by destroying dynamical localization. With a suitably chosen non-zero
bias ($\delta p \neq 0$), we find that the noisy field can destroy
dynamical localization yielding a delocalized phase. On the other
hand, there is a way to tune the noise such that it induces dynamical
localization in an otherwise delocalized phase of the clean system. A
study of the entanglement entropy in the many-body setting provides
useful signatures for all these effects. We corroborate our analytical
calculation with an exact numerical approach by a study of the
probability propagator and the dynamics of entanglement entropy.

We have organized our work as follows. In Section~\eqref{Model
  hamiltonian}, we introduce the model Hamiltonian along with the
unitary transformation to obtain an effective Hamiltonian. In
Section~\eqref{Observables}, we derive the expression for
the probability propagator $\tilde{\mathcal{P}}_{n}\left(t\right)$ and provide a
brief discussion. The clean limit of the problem is discussed in
Section \eqref{Clean Limit Case}, while Section \eqref{Effect of
  Stochastic Noise} is devoted to the case of stochastic
noise. Finally, we have summarized our main findings in Section
\eqref{Summary and Conclusion}.

\section{MODEL HAMILTONIAN}\label{Model hamiltonian}
We consider the dynamics of a single-particle moving in a one dimensional tight binding chain under the influence of a time-dependent field comprising of a square wave electric field and time dependent random telegraph noise. The model Hamiltonian can be written as:
\begin{eqnarray}
\label{eqn:eq1a}
H&=& -\frac{\Delta}{4}\left(\sum_{n=-\infty}^{\infty}|n+1\rangle \langle n| + |n\rangle \langle n+1| \right)+\nonumber\\
&&\mathcal{F}\left(t\right)\sum_{n=-\infty}^{\infty}n | n\rangle \langle n|,
\end{eqnarray}
where $|n\rangle$ is a Wannier state localized at site $n$ and $\Delta$ is the hopping strength. $\mathcal{F}\left(t\right)$ is the time dependent field: $\mathcal{F}\left(t\right)= F\left(t\right)+\xi\left(t\right)$, where $F\left(t\right)$ is the square wave electric field and $\xi\left(t\right)$ is a time-dependent stochastic telegraph noise. The square wave drive can be expressed as 
\begin{eqnarray}
F\left(t\right)=  \left\{\begin{array}{cc} +A,  & 0\leq t \leq T/2  \\
- A,  & T/2<t \leq T \end{array}\right., \label{eqn:eq3} 
\end{eqnarray}
where $T$ is the time period of the drive. We work in units where $\hbar=e=1$ and the lattice constant too is unity. 

To study the dynamical evolution, we start by defining the unitary operators~\cite{Hartmann_2004}:
\begin{eqnarray}
\label{eqn:eq2a}
\hat{K}&=&\sum_{n=-\infty}^{n=\infty}|n\rangle \langle n+1| \nonumber\\
\hat{K}^{\dagger}&=&\sum_{n=-\infty}^{n=\infty}|n+1\rangle \langle n|\nonumber\\
\hat{N}&=&\sum_{n=-\infty}^{\infty}n|n\rangle \langle n|.
\end{eqnarray}
These operators are diagonal in the quasi-momentum basis $k$:
\begin{eqnarray}
\langle k^{\prime}| \hat{K} |k \rangle &=& e^{ik}\delta\left(k-k^{\prime}\right), \nonumber\\
\langle k^{\prime}| \hat{K^{\dagger}} |k\rangle &=& e^{-ik}\delta\left(k-k^{\prime}\right),
\end{eqnarray}
and follow the commutation relations:
\begin{eqnarray}
\label{eqn:eq3a}
\left[\hat{K}, \hat{N}\right]=\hat{K}, & \left[\hat{K}^{\dagger}, \hat{N}\right]=-\hat{K}^{\dagger}, &\left[\hat{K}, \hat{K}^{\dagger}\right]=0.
\end{eqnarray}  
In terms of these unitary operators, the Hamiltonian~(\ref{eqn:eq1a}) can be written as $\hat{H} = V_{+} + H_{0}\left(t\right)$, where 
\begin{eqnarray}
\label{eqn:eq5a}
 V_{+}=-\frac{\Delta}{4}\left(\hat{K}+\hat{K}^{\dagger}\right),\ \ \ H_{0}\left(t\right)&=&\mathcal{F}\left(t\right)\hat{N}.
\end{eqnarray}
The equation of motion for the density matrix in the Heisenberg picture is
\begin{eqnarray}
\frac{d\rho}{dt}&=&-i\left[H\left(t\right), \rho\right].
\end{eqnarray}
After the unitary transformation, we can write the density matrix $\rho\left(t\right)$ as
\begin{eqnarray}
\label{eqn:eq6a}
\tilde{\rho}\left(t\right)&=& e^{i\int_{0}^{t}H_{0}\left(t^{\prime}\right)dt^{\prime}}\rho\left(t\right)e^{-i\int_{0}^{t}H_{0}\left(t^{\prime}\right)dt^{\prime}}.
\end{eqnarray}
Following the procedure of Bhakuni et al~\cite{PhysRevB.99.155149}, we obtain the density operator in the momentum basis as
\begin{eqnarray}
\label{eqn:eq12a}
\langle k\mid \tilde{\rho}\left(t\right)\mid k^{\prime}\rangle&=& e^{-i\int_{0}^{t}\tilde{V}_{+k}\left(t^{\prime}\right)dt^{\prime}}\langle k \mid 0\rangle \langle 0\mid  k^{\prime}\rangle e^{i\int_{0}^{t}\tilde{V}_{+k}\left(t^{\prime}\right)dt^{\prime}},\nonumber\\
\end{eqnarray}
where, $\tilde{V}_{+k}\left(t^{\prime}\right)=-\frac{\Delta}{4}\left[e^{i\left(k+\eta\left(t^{\prime}\right)\right)}+ e^{-i\left(k+\eta\left(t^{\prime}\right)\right)}\right]$.
This suggests that the dynamical evolution of the system is governed by an effective Hamiltonian $\tilde{V}_{+k}$ where the time-dependence comes only as a phase factor. Furthermore the effective Hamiltonian respects translation invariance and hence allows us to calculate the dynamical evolution of the observables analytically as described ahead.    

\section{Observables} \label{Observables}

In order to study the dynamics of single particle, we derive an expression for probability propagator. The probability propagator $\mathcal{P}_{n}\left(t\right)$ is a
measure of the probability of finding the particle at site $n$ at a
time $t$. Here, we consider an initial state where the particle is
localized at the central site ($n=0$). The probability propagator can
be defined as~\cite{PhysRevB.99.155149}:

\begin{eqnarray}
\label{eqn:eq14a}
\mathcal{P}_{n}\left(t\right)&=& \int_{-\pi}^{\pi} \int_{-\pi}^{\pi} dk\ dk^{\prime} \langle n \mid k\rangle \langle k \mid \tilde{\rho}\left(t\right)\mid k^{\prime}\rangle \langle k^{\prime}\mid n \rangle.
\end{eqnarray}
Using  Eq.(\ref{eqn:eq12a}), we can write
\begin{eqnarray}
\label{eqn:eq15a}
 \mathcal{P}_{n}\left(t\right)&=& \left(\frac{1}{2\pi}\right)^{2}\int_{-\pi}^{\pi} dk \int_{-\pi}^{\pi} dk^{\prime} e^{-i\left(k-k^{\prime}\right)n}\times\nonumber\\
&& e^{-i\int_{0}^{t}dt^{\prime}\left[\tilde{V}_{+k}\left(t^{\prime}\right)-\tilde{V}_{+k^{\prime}}\left(t^{\prime}\right)\right]}.
\end{eqnarray}

In our case, the time dependent field is a combination of square wave pulse and time dependent telegraph noise \cite{doi:10.1063/1.1150519}. The noise consists of random jumps between two levels $\pm \mu$. By denoting $\sigma$ and $\tau$ to be the rate of switching from level $+\mu$ to $-\mu$ and $-\mu$ to $+\mu$,  respectively, the probability of being at any time in state $+\mu$ can be defined as: $p_{+}=\frac{\tau}{\lambda}$, whereas the probability of being in state $-\mu$ is: $p_{-}=\frac{\sigma}{\lambda}$ where we define $\lambda= \sigma+\tau$.

For such noise, the overall field can be expressed as a sum of $2\times 2$ matrices~\cite{PhysRev.174.351}:
\begin{eqnarray}
\label{eqn:eq17}
i\eta \left(t\right)=i\int_{0}^{t}F(t^{\prime})dt^{\prime}\mathcal{I}+it\mu \sigma_{z}+ t W.
\end{eqnarray}
Here, $W$ is the relaxation matrix defined as~\cite{PhysRevB.82.245417,dattagupta2012relaxation}
\begin{equation}
\label{eqn:eq2}
W=
\begin{bmatrix}
-p_{-} & p_{+}\\
p_{-} & -p_{+}\\
\end{bmatrix}
=\lambda
\begin{bmatrix}
-\frac{\sigma}{\tau+\sigma} & \frac{\tau}{\tau+\sigma}\\
\frac{\sigma}{\tau+\sigma}  & -\frac{\tau}{\tau+\sigma}
\end{bmatrix}
\end{equation}
 and $\mathcal{I}$ is the identity matrix. From Eq.(\ref{eqn:eq2}),
 the relaxation matrix $W$ can be expressed as a linear combination of
 the identity matrix $\mathcal{I}$ and Pauli matrices
 $\sigma_{i}^{\prime}s$ having components $h_{0}=-h_{1}=-\gamma,
 h_{2}=i\delta p \gamma, h_{3}=\left(\gamma\delta p+i\mu\right)$,
 where $\gamma =\frac{\sigma +\tau}{2}=\frac{\lambda}{2}$, $\delta p$
 is the difference between the probabilities and equals
 $\left(p_{+}-p_{-}\right)$.

To proceed with the calculation of the probability propagator, we
express the exponential of Eq.\eqref{eqn:eq17} in compact form as :
\begin{eqnarray}
\label{eqn:eq7}
e^{i\eta\left(t^{\prime}\right)}&=&e^{i\sum_{l=1}^{\infty}\alpha_{l}\left(1-\cos \left( \omega \left(2l-1\right) t^{\prime}\right)\right)}e^{-\gamma t^{\prime}}\frac{1}{2}\left[e^{\nu t^{\prime}}\left(1+\hat{h}.\vec{\sigma}\right)+\right.\nonumber\\
&&\left.e^{-\nu t^{\prime}}\left(1-\hat{h}.\vec{\sigma}\right)\right],
\end{eqnarray}
where $\alpha_{l}
=\frac{4AT}{2\left(2l-1\right)^{2}\pi^{2}},\ \omega=\frac{2\pi}{T}$,
$l$ is an integer and $\nu=\sqrt{h_{1}^{2}+h_{2}^{2}+h_3^{2}}$. Here,
we have exploited the identity satisfied by Pauli matrices:
$e^{i\left(\vec{a}\vec{\sigma}\right)}=\left(\mathcal{I}\cos(|
\textbf{a}|)+i\left(\hat{a}.\vec{\sigma}\right)\sin\left(|
\textbf{a}|\right)\right)$.

Expressing the square wave in terms of its Fourier series components, Eq.(\ref{eqn:eq7}) can be written as:
\begin{eqnarray}
\label{eqn:eq8}
e^{i\eta\left(t^{\prime}\right)} &=& \left\{\begin{array}{cc}\frac{1}{2}e^{\left(iAt^{\prime}-iAnT\right)} e^{-\gamma t^{\prime}}\left[e^{\nu t^{\prime}}\left(1+\hat{h}.\vec{\sigma}\right)+\right.\\
\left.e^{-\nu t^{\prime}}\left(1-\hat{h}.\vec{\sigma}\right)\right],\\
\left(2n\pi \leq\omega t^{\prime}\leq \left(2n+1\right)\pi\right); \nonumber \\
\frac{1}{2}e^{-iAt^{\prime}+iA\left(n+1\right)T}e^{-\gamma t^{\prime}}\left[e^{\nu t^{\prime}}\left(1+\hat{h}.\sigma\right)+\right.\nonumber \\
\left.e^{-\nu t^{\prime}}\left(1-\hat{h}.\vec{\sigma}\right)\right],\nonumber \\
\left(\left(2n+1\right)\pi <\omega t^{\prime}\leq 2\left(n+1\right)\pi \right). \end{array}\right.\\
\end{eqnarray}

After detailed calculations (See Appendix \ref{app:subsec}), we obtain an expression for the probability propagator:
\begin{eqnarray}
\label{eqn:eq10}
\mathcal{P}_{n}\left(t\right)=
\left(\frac{1}{2\pi}\right)^{2}\int_{-\pi}^{\pi}dk\int_{-\pi}^{\pi}dk^{\prime}e^{-i\left(k-k^{\prime}\right)n}e^{ig_{0}\left(t\right)}\nonumber\\
\times \left(\mathcal{I}\cos\left(|\textbf{H}|\right)+i\left(\hat{\textbf{H}}.\vec{\sigma}\right)\sin\left(|\textbf{H}|\right)\right),
\end{eqnarray}
where we have defined $H_{x}=g_{1}(t),
H_{y}=g_{2}\left(t\right)=i\delta pg_{1}\left(t\right),
H_{z}=g_{3}\left(t\right)=\delta
pg_{1}\left(t\right)+\beta\left(t\right)$ and $g_{0}(t), g_{1}(t),
g_{2}(t)$ and $g_{3}(t)$ are defined in Appendix~\eqref{app:subsec}.
The final expression for average probability can be obtained by
calculating $ \tilde{\mathcal{P}}_{n}\left(t\right)=\sum_{ab}p_{a}
\langle b|\mathcal{P}_{n}\left(t\right)|a\rangle$. This requires an
average of various Pauli matrices with respect to the available stochastic states $|+\rangle = \begin{bmatrix} 1 \\ 0\end{bmatrix}$
  and $|-\rangle= \begin{bmatrix} 0 \\ 1 \end{bmatrix}$. In order to study the dynamics of the system, we
use two observables: probability propagator to explore
the single-particle dynamics and entanglement entropy
to explore the features of a noninteracting many-fermion
quantum system.
 This results
  in $\langle \sigma_{x}\rangle=1, \langle \sigma_{y}\rangle= -i\delta
  p, \langle \sigma_{z}\rangle =\delta p$, and $\langle
  \mathcal{I}\rangle=1$. With this simplification, we get the final
  expression of the probability propagator as
\begin{eqnarray}
\label{eqn:eq11}
\tilde{\mathcal{P}}_{n}\left(t\right)=\left(\frac{1}{2\pi}\right)^{2}\int_{-\pi}^{\pi}dk\int_{-\pi}^{\pi}dk^{\prime}e^{-i\left(k-k^{\prime}\right)n}e^{ig_{0}\left(t\right)}\times \qquad\qquad \nonumber\\
\left(\cos\left(|\textbf{H}|\right)+i\frac{g_{1}\left(t\right)}{|\textbf{H}|}\sin\left(|\textbf{H}|\right)+i\sin\left(|\textbf{H}|\right)\frac{\delta p \beta\left(t\right)}{|\textbf{H}|}\right).\qquad
\end{eqnarray}
Hence, we manage to obtain an exact expression for the probability
propagator valid for both ac and dc electric fields in the presence of
a telegraph noise.

To explore the effect of noisy drive in the (non-interacting) many
body state, we have also calculated von-Neumann entanglement
entropy~(\ref{eqn:eq12}) between two halves of the chain. For the
dynamical evolution, we take the initial state $|\Psi_{in}\rangle$
where all the particles are localized to the left side of the chain
\begin{equation}
\label{eqn:eq16}
|\Psi_{in}\rangle=c_1^{\dagger}c_2^{\dagger}....c_{N/2}^{\dagger}|0\rangle,
\end{equation}
where $c_{i}^{\dagger}$ is the creation operator at site $i$. 
To study the dynamics of entanglement entropy, we form a time-dependent correlation matrix~\cite{Peschel_2003,PhysRevB.97.125116} as 
\begin{eqnarray}
\label{eqn:eq21}
C_{mn}\left(t\right)&=&\langle \Psi_{in}\left(t\right)|c_{m}^{\dagger}c_{n}|\Psi_{in}\left(t\right)\rangle\nonumber\\
&&=\langle \Psi_{in}\left(0\right)|c_{m}^{\dagger}\left(t\right)c_{n}\left(t\right)|\Psi_{in}\left(0\right)\rangle,
\end{eqnarray}
which can be simplified to
\begin{eqnarray}
\label{eqn:eq22}
C\left(t\right)=U^{\dagger}\left(t\right)C\left(0\right)U\left(t\right),
\end{eqnarray}
where
$U_{jk}\left(t\right)=\sum_{n}D_{jn}^{*}\exp\left(-i\epsilon_{n}t\right)D_{nk}$
and the matrix $D$ diagonalizes the final Hamiltonian. The detailed
calculation to obtain entanglement entropy from the correlation matrix
is given in Appendix~\ref{app1:EE}. Diagonalizing the time-dependent
correlation matrix and invoking Eq.\eqref{eqn:eq15}, we can study the
dynamics of entanglement entropy.

\begin{figure*}[t]
	\includegraphics[scale=0.35]{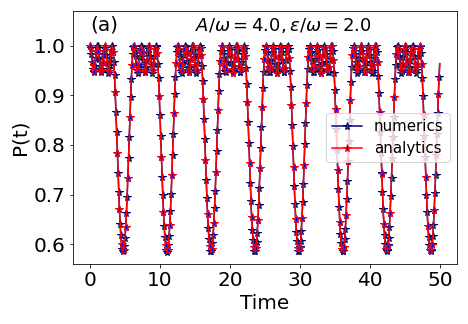}
	\includegraphics[scale=0.35]{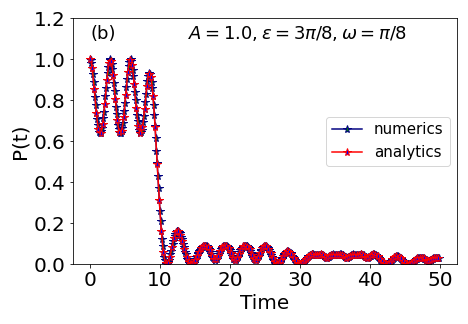}
	\includegraphics[scale=0.35]{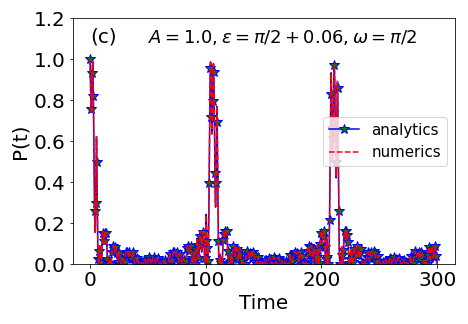}
	\includegraphics[scale=0.35]{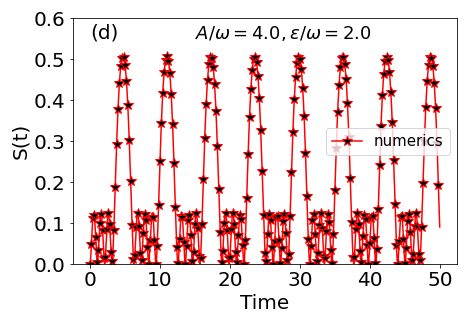}
	\includegraphics[scale=0.35]{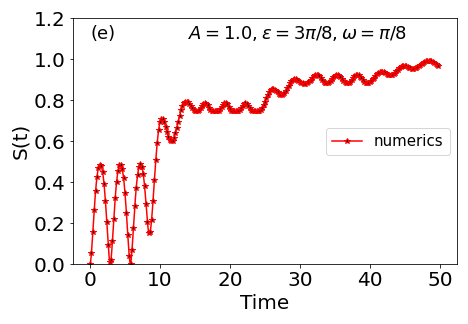}
	\includegraphics[scale=0.35]{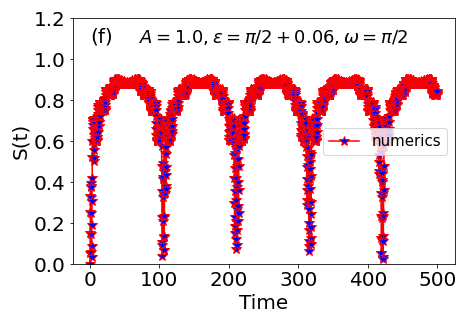}
	\caption{(a-c) Return probability for AC+DC driven system. (a)
          Dynamical localization for resonantly tuned DC
          drive ($\epsilon=n^{\prime}\omega$,with $n^{\prime}$ even).
          (b) Coherent destruction of Wannier-Stark localization at
          values away from the dynamical localization conditions. (c) Super Bloch
          oscillations at slightly detuned DC drive
          ($\epsilon=\left(n^{\prime}+\delta\right)\omega$). (d-f)
          Entanglement entropy for AC+DC driven system. (d)
          Periodicity signifies dynamical localization for resonantly
          tuned DC drive ($\epsilon=n^{\prime}\omega $). (e) Unbounded
          growth signifies delocalization at values away from dynamical
          localization conditions. (f) Periodic oscillations
          correspond to super Bloch oscillations. The other parameters
          are $L=200, \Delta=2.0$.  }
	\label{fig:fig1}
\end{figure*}

\section{The driven clean system}\label{Clean Limit Case}
Having obtained the exact expression for the probability propagator,
we now proceed to discuss different forms of the electric field, and
the various phenomena associated with them.  We first consider the
case where the noise is absent $\left(\mu =0\right)$ and the system is
driven by a time-periodic square wave pulse. In this limit
$\nu\rightarrow \gamma$ and $\beta\rightarrow 0$, and
Eq.(\ref{eqn:eq11}) becomes
\begin{eqnarray}
\label{eqn:eq23}
\tilde{\mathcal{P}}_{n}\left(t\right)  =  \left(\frac{1}{2\pi}\right)^{2}\int_{-\pi}^{\pi}dk\int_{-\pi}^{\pi}dk^{\prime}e^{-i\left(k-k^{\prime}\right)n}e^{ig_{0}\left(t\right)}e^{ig_{1}\left(t\right)}.\qquad
\end{eqnarray}
where $g_{0}(t)$ and $g_1(t)$ are defined in the Appendix (Eq.\eqref{eqn:eqapph} and \eqref{eqn:eqappi}). For the case of a pure noise-free square wave drive (Eq.~\ref{eqn:eq3}), we have (as shown in the Appendix):
\begin{eqnarray}
\label{eqn:eq24}
\lim_{\mu\rightarrow 0}\left(g_{0}\left(t\right)+g_{1}\left(t\right)\right)& = &\frac{\Delta}{2A}\left\lbrace \frac{\left(t-\tau\right)}{\pi/\omega}\left(\sin\left(k+\pi A/\omega\right)\right.\right.\nonumber\\
&&\left.\left.
-\sin k\right)+\left(\sin\left(k+A\tau\right)-\sin k\right)-\right.\nonumber\\
&&\left.\frac{\left(t-\tau\right)\omega}{\pi}\left(\sin\left(k^{\prime}+\pi A/\omega\right)-\sin k^{\prime}\right)\right.\nonumber\\
&&\left.-\left(\sin\left(k^{\prime}+A\tau\right)-\sin k^{\prime}\right)\right\rbrace,
\end{eqnarray}
where $t=mT+\tau, \quad\left(0< \tau < T/2\right)$ and $m$ is a
positive integer. In the limit $\omega\to 0$, we have a static field, and it is evident that only the second and fourth terms in Eq(\ref{eqn:eq24}) survive, and yield the familiar Bloch oscillations. For nonzero $\omega$, we observe that Eq(\ref{eqn:eq24}) becomes a periodic function with period
$\frac{2\pi}{A}$ if the ratio $\frac{A}{\omega}$ is tuned to an even integer. This
periodicity corresponds to dynamical localization
at these specific ratios of the amplitude and frequency as reported
previously~\cite{SciPostPhys.3.4.029, PhysRevA.79.013611}. On the
other hand, going away from these special points, for $A/\omega=$odd
integer, we have
\begin{eqnarray}
\label{eqn:eq25p}
\lim_{\mu\rightarrow 0}\left(g_{0}\left(t\right)+g_{1}\left(t\right)\right)& = & \left\lbrace \frac{\Delta\zeta}{A}\left(\sin\left(-C+k\right)\right)\right.\nonumber\\
&&\left.-\frac{\Delta\zeta}{A}\left(\sin\left(-C+k^{\prime}\right)\right)\right\rbrace,
\end{eqnarray}
where $\zeta= \sqrt{\frac{\left(t-\tau\right)^{2}}{\pi^{2}/\omega^{2}}+\sin ^{2}A\tau}, \cos C= \frac{\left(t-\tau\right)}{\zeta \pi/\omega}, \sin C= \frac{\sin A\tau}{\zeta}$. This gives a closed-form expression for the probability propagator: 
\begin{eqnarray}
\label{eqn:eq26p}
\tilde{\mathcal{P}}_{n}\left(t\right)  &=& \left(J_{0}\left(\Delta\zeta/A\right)\right)^{2},
\end{eqnarray}
which suggests a decaying behavior in time and hence signifies the delocalization of an initially localized wave-packet.  

Another interesting case arises when the electric field has both ac
and dc parts. Numerical and semi-classical
studies~\cite{PhysRevB.102.024201,PhysRevB.98.045408} on such combined
ac and dc electric fields have revealed several interesting phenomena
such as coherent destruction of Wannier-Stark localization, dynamical
localization, and super-Bloch oscillations. Here, we discuss these
phenomena in the backdrop of our exact result for the probability
propagator. When the system is subjected to a square wave drive with
amplitude $A$ and frequency $\omega$ along with a uniform dc field
$\epsilon$ i.e. $\mathcal{F}\left(t\right)=F\left(t\right)+\epsilon$,
~Eq.\eqref{eqn:eq24} gets modified to:
\begin{widetext}
\begin{eqnarray}
\label{eqn:eq27an}
\lim_{\mu\rightarrow 0}\left(g_{0}\left(t\right)+g_{1}\left(t\right)\right) & =&\lim_{\mu\rightarrow 0}\frac{\Delta}{4}\left[\left(z-z^{\prime}\right)\left\lbrace\left( \frac{1-e^{-\left(-i\epsilon-iA\right)\pi/\omega}}{\left(-i\epsilon-iA\right)}+\frac{\left(1-e^{-\left(-i\epsilon+iA\right)\pi/\omega}\right)e^{-\left(-i\epsilon+iA\right)\pi/\omega}e^{iA.2\pi/\omega}}{\left(-i\epsilon+iA\right)}\right)\right.\right.\nonumber\\
&&\left.\left. \left(\frac{1-e^{-\left(\gamma-\nu-i\epsilon\right)2m\pi/\omega}}{1-e^{-\left(\gamma-\nu-i\epsilon\right)2\pi/\omega}}\right)+
\frac{e^{i\epsilon \frac{2m\pi}{\omega}}\left(1-e^{-\left(-i\epsilon-iA\right)\left(\tau\right)}\right)}{\left(-i\epsilon-iA\right)}\right\rbrace+c.c.\right]
\end{eqnarray}
where $z=e^{ik}, z^{\prime}=e^{i k^{\prime}}$, which in turn leads to
\begin{eqnarray}
\label{eqn:eq25a}
\lim_{\mu\rightarrow 0}\left(g_{0}\left(t\right)+g_{1}\left(t\right)\right)& = &\frac{\Delta}{2}\left\lbrace \frac{\left(t-\tau\right)}{\left(A+\epsilon\right)\frac{2\pi}{\omega}}\left[\left(\sin\left(k+\pi \left(A+\epsilon\right)/\omega\right)
-\sin k\right)-\left(\sin\left(k^{\prime}+\pi \left(A+\epsilon\right)/\omega\right)
-\sin k^{\prime}\right)\right]\right.\nonumber\\
&&\left.+\frac{\left(t-\tau\right)}{\left(A-\epsilon\right)\frac{2\pi}{\omega}}\left[\left(\sin\left(k+\pi \left(A+\epsilon\right)/\omega\right)
-\sin \left(k+\frac{2\epsilon\pi}{\omega}\right)\right)-\left(\sin\left(k^{\prime}+\pi \left(A+\epsilon\right)/\omega\right)
-\right.\right.\right.\nonumber\\
&&\left.\left.\left.
\sin \left(k^{\prime}+\frac{2\epsilon\pi}{\omega}\right)\right)\right]+\frac{1}{\left(A+\epsilon\right)}\left[\sin\left(k+\left(A+\epsilon\right)\tau+\frac{2m\epsilon\pi}{\omega}\right)-\sin\left(k+\frac{2m\epsilon\pi}{\omega}\right)\right]\right.\nonumber\\
&&\left.-\frac{1}{\left(A+\epsilon\right)}\left[\sin\left(k^{\prime}+\left(A+\epsilon\right)\tau+\frac{2m\epsilon\pi}{\omega}\right)-\sin\left(k^{\prime}+\frac{2m\epsilon\pi}{\omega}\right)\right]
\right\rbrace.
\end{eqnarray}
\end{widetext}

Eq.\eqref{eqn:eq25a} suggests that certain special ratios of the
static field with the frequency of the square wave drive may yield
interesting dynamical phenomena.  For $A/\omega=2n$, we see periodic
behavior if the static field is tuned at $\epsilon/\omega=2n^{\prime},
\left(n\neq n^{\prime}\right)$. This corresponds to dynamical
localization as shown in Fig.\eqref{fig:fig1}(a,d), where the
initially localized wave-packet returns to its initial state after a
driving period and both the probability and the entanglement entropy
oscillate in time. If the static field is resonantly tuned with
$\epsilon/\omega=2n^{\prime}$, but the ratio $A/\omega$ is set to be
something other than an even integer, the driving leads to band
formation~\cite{PhysRevB.102.024201} and destroys the localization set
up by the static field. This coherent destruction of Wannier Stark
localization is shown in Fig.\eqref{fig:fig1}(b,e) where the
probability decays in time and the entropy shows an unbounded growth
despite the presence of the static field. Similarly, if the static
field is resonantly tuned at $\epsilon/\omega=2n^{\prime}+1,
\left(n\neq n^{\prime}\right)$, dynamical localization can again be
observed if $A/\omega=2n+1$ is also an odd integer. However, while
maintaining the resonance condition, if the static field is not tuned
in this precise manner, we once again observe coherent destruction of
Wannier-Stark localization.

Finally, when the static electric field is slightly detuned from the
resonance condition such that $\epsilon=\left(n^{\prime}+\delta
\right)\omega$, the phase factor in Eq.~\eqref{eqn:eq25a} acquires an
extra term $\delta \omega t$ and it becomes
\begin{widetext}
\begin{eqnarray}
\label{eqn:eq26a}
\lim_{\mu\rightarrow 0}\left(g_{0}\left(t\right)+g_{1}\left(t\right)\right)& = &\frac{\Delta}{2}\left[\frac{1}{\left(A+\epsilon\right)\sin(\delta\pi)}\left\lbrace \sin\left(k^{\prime}+(m-1)\delta\pi\right)-\sin\left(k+(m-1)\delta\pi\right)\right\rbrace\sin(m\delta\pi)+\frac{1}{\left(A-\epsilon\right)\sin(\delta\pi)}\right.\nonumber\\
&&\left. \left\lbrace \sin\left(k^{\prime}+\delta\pi+m\delta \pi\right)-\sin\left(k+\delta\pi+m\delta \pi\right)\right\rbrace\sin(m\delta\pi)+\frac{2A}{\left(A^{2}-\epsilon^{2}\right)\sin(\delta\pi)}\right.\nonumber\\
&&\left.\left\lbrace \sin\left(k+\frac{A\pi}{\omega}+n^{\prime}\pi+m\delta\pi\right)-\sin\left(k^{\prime}+\frac{A\pi}{\omega}+n^{\prime}\pi+m\delta\pi\right)\right\rbrace \sin\left(m\pi\delta\right)+\frac{2}{\left(A+\epsilon\right)}\right.\nonumber\\
&&\left.\left\lbrace\sin\left(k+(n^{\prime}+\delta)\omega t+A\tau\right)-\sin\left(k+m\delta\omega T\right)-\sin\left(k^{\prime}+(n^{\prime}+\delta)\omega t+A\tau\right)+\sin\left(k^{\prime}+m\delta\omega T\right)\right\rbrace\right].\nonumber\\
\end{eqnarray}
\end{widetext}

\begin{figure*}[t]
	\includegraphics[scale=0.265]{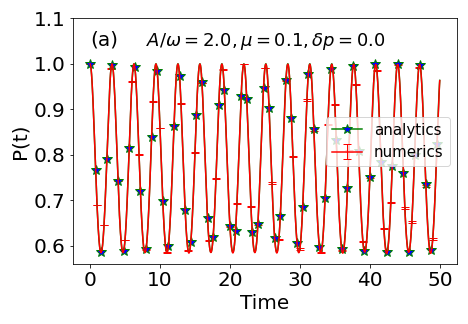}
	\includegraphics[scale=0.265]{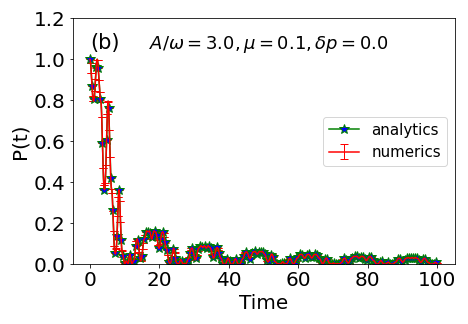}
	\includegraphics[scale=0.265]{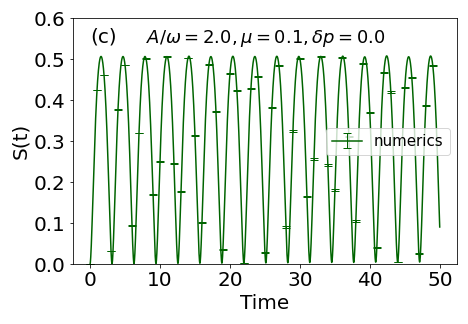}
	\includegraphics[scale=0.265]{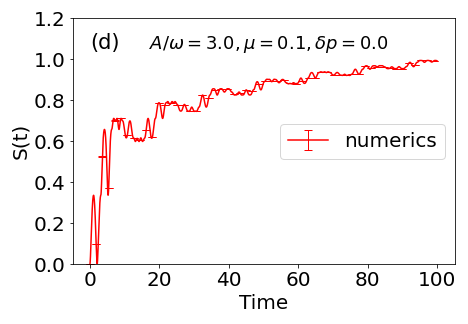}
	\caption{Data for fluctuating square wave driven system. (a)
          shows the periodic behavior of return probability at
          dynamical localization point. (b) shows decaying behavior of
          probability at points away from dynamical localization. (c)
          shows periodic behavior of entanglement entropy at dynamical
          localization point and (d) shows unbounded growth of
          entanglement entropy at points away from dynamical
          localization conditions. We present data for the zero bias case
          $\left(\delta p=0.0\right)$ in the rapid relaxation regime
          $\left(\sigma=100, \tau=100\right)$ with hopping strength
          $\Delta=2.0$.  We have averaged over 200 realizations of
          disorder for a system of size $L=200$.}
	\label{fig:fig2}
\end{figure*}
The system exhibits oscillatory behaviour similar to Bloch
oscillations. Analogous to the case of the static field driven system,
these oscillations are termed as super Bloch oscillations and the
frequency of these oscillations is directly proportional to $\delta
\omega$. This phenomenon has been shown in Fig.\eqref{fig:fig1}(c,f)
where return probability and entanglement entropy exhibit periodic
behaviour with the frequency given by the offset.

\section{Effect of Stochastic Noise}\label{Effect of Stochastic Noise}
In this section, we will focus on how the presence of time dependent
fluctuations affect the dynamical localization in the system. We
present an analytical expression for the probability propagator and
also test its validity with the aid of an exact numerical approach. To
simulate the telegraph noise and the dynamical protocol, we follow
Bhakuni et al~\cite{PhysRevB.99.155149} and average the observables over many noise
trajectories. We will restrict ourselves to the case of zero static
field and discuss the effects of the inclusion of a noisy field. We
consider two different cases: one where the two levels of the noise
are equiprobable ($\delta p =0$) and the other where one level is more
probable ($\delta p \neq 0$). While the expression for the probability
propagator is general, we will restrict ourselves to the rapid
relaxation regime, in order to obtain approximate expressions that are
effective and simple. In this limit $\gamma >> \mu, A$ and
$g_{1}\left(t\right)^{2}>>\beta\left(t\right)^{2}$. With these
approximations, we can expand $|\textbf{H}|$ as
\begin{eqnarray}
\label{eqn:eq26}
|\textbf{H}| &= &\sqrt{g_{1}\left(t\right)^{2}+\beta\left(t\right)^{2}+2\delta pg_{1}\left(t\right)\beta\left(t\right)}\nonumber\\
&&\approx   g_{1}\left(t\right)\left(1+\frac{\beta\left(t\right)^{2}}{2g_{1}\left(t\right)^{2}}+\frac{\delta p\beta\left(t\right)}{g_{1}\left(t\right)}\right),
\end{eqnarray}
and the expression which appears in the integrand of Eq.~(\ref{eqn:eq11}) can be approximated as

\begin{align}
\label{eqn:eq27}
e^{ig_{0}\left(t\right)}&\left(\cos\left(|\textbf{H}|\right)+i\frac{\sin(|\textbf{H}|)}{|\textbf{H}|}\Big[g_{1}\left(t\right)+\delta p \beta\left(t\right)\Big]\right)\nonumber\\
&\approx \exp\Big(i(g_{0}\left(t\right)+|\textbf{H}|)\Big)\qquad\qquad\nonumber\\
&\approx \exp{(ig_{0}(t)+ig_{1}(t)+i\delta p\beta(t)+i\frac{\beta^{2}(t)}{2g_{1}(t)}}).\nonumber\\
\end{align}

First, we consider the case of zero bias $\left(\delta p =0\right)$
with zero external drive. In this limit
$\nu=\sqrt{\gamma^{2}-\mu^{2}}\approx
\left(\gamma-\frac{\mu^{2}}{2\gamma}\right)$. When the field is zero
$\left(A=0\right)$, Eq.\eqref{eqn:eq27} is modified to
Eq.\eqref{eqn:app2c} (see Appendix), which in the long-time limit,
futher simplifies to
\begin{eqnarray}
\label{eqn:eq28a}
e^{ig_{0}\left(t\right)+ig_{1}\left(t\right)}e^{i\frac{\beta^{2}\left(t\right)}{2g_{1}\left(t\right)}}& \approx &e^{i\frac{\Delta_{\textsf{eff}}}{4}t\left(\cos k-\cos k^{'}\right)},
\end{eqnarray}
where
$\Delta_{\textsf{eff}}=\Delta\left(1+\frac{1}{2}\left(\frac{\mu}{\gamma}\right)^{2}\frac{\left(\sin
  k-\sin k^{\prime}\right)^{2}}{\left(\cos k-\cos
  k^{\prime}\right)^{2}}\right)$.  This expression shows that in the
zero field limit, the dynamics is governed by a renormalized hopping
parameter $\Delta_{\textsf{eff}}$ for rapidly fluctuating noise, thus
recovering an earlier result~\cite{PhysRevB.99.155149}.

Now, we consider the square wave driven system $\left(A\neq 0\right)$.
In this case, we can ignore
$e^{i\frac{\beta^{2}\left(t\right)}{2g_{1}\left(t\right)}}$ and
Eq.\eqref{eqn:eq27} is approximated to Eq.\eqref{eqn:app2e}. In the
rapid relaxation limit, for $A/\omega=2n,$ Eq.\eqref{eqn:app2e} further simplifies to:
\begin{align}
\label{eqn:eq29}
g_{0}\left(t\right)+g_{1}\left(t\right) &\approx \frac{-\Delta e^{-\frac{\mu^{2}}{2\gamma}t}}{2A}\Big[e^{\frac{\mu^{2}\tau}{2\gamma}}\left(\sin k-\sin k^{\prime}\right)\nonumber\\&- \left(\sin \left(k+A\tau\right)-\sin \left(k^{\prime}+A\tau\right)\right)\Big].
\end{align}

It is clear from Eq.(\ref{eqn:eq29}) that
$g_{0}\left(t\right)+g_{1}\left(t\right)$ will exhibit periodic
behavior (with period $\frac{2\pi}{A}$) only for small values
of noise ($\mu<<\gamma$), whereas for large values of noise and in the long time
limit, the periodic oscillations damp out exponentially. The dynamics
of return probability and entanglement entropy are plotted in
Fig.(\ref{fig:fig2}). At the dynamical localization point, we see the
oscillatory nature of the probability propagator
(Fig.~\ref{fig:fig2}(a)) and the entropy
(Fig.~\ref{fig:fig2}(c)). However these oscillations are bound to
decay on much longer time scales. Similarly, when the parameters are
tuned away from the dynamical localization point, we see a decaying
behavior of probability and an unbounded growth of the entanglement
entropy as depicted in Fig.~\eqref{fig:fig2}(b) and
Fig.~\eqref{fig:fig2}(d) respectively.

We next consider the case where one level of the telegraph noise is
more probable than the other i.e. $\delta p \neq 0$. In this case, we
can approximate $\nu$ as:
$\nu=\sqrt{\gamma^{2}-\mu^{2}+2i\mu\gamma\delta p}\approx
\left(\gamma-\frac{\mu^{2}}{2\gamma}+i\mu\delta p\right)$. For $A/\omega=2n$, with this
approximation and further simplifications (via Eq.~\eqref{eqn:app2i}) we arrive at:
 \begin{widetext}
	\begin{eqnarray}
	\label{eqn:eq30}
	g_{0}(t)+g_{1}(t)+\delta p\beta(t)& \approx & \frac{-\Delta e^{-\frac{\mu^{2}}{2\gamma}t}}{2(\mu\delta p+A)}\Big[e^{\frac{\mu^{2}}{2\gamma}\tau} \left\lbrace \sin(k+\mu\delta p(t-\tau))-\sin(k^{\prime}+\mu\delta p(t-\tau))\right\rbrace -\nonumber\\
	&&
\left\lbrace \sin \left(k+\mu \delta p t+A\tau\right)-
\sin\left(k^{\prime}+\mu\delta p t+A\tau\right)\right\rbrace\Big]. 
	\end{eqnarray}
\end{widetext}

\begin{figure}[t]
	\includegraphics[scale=0.25]{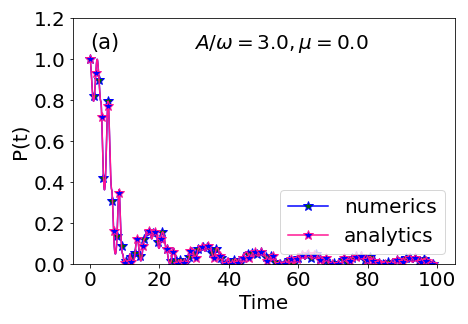}
	\includegraphics[scale=0.25]{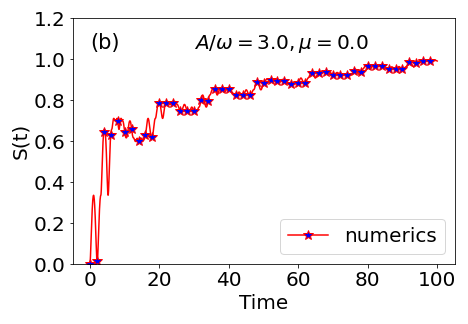}
	\includegraphics[scale=0.25]{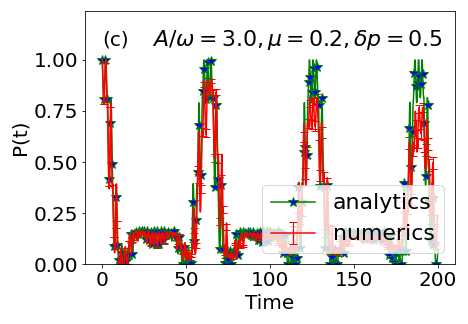}
	\includegraphics[scale=0.25]{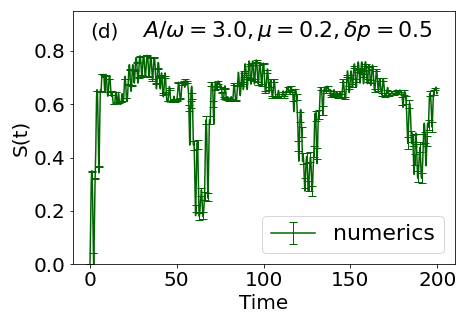}
	\caption{Return probability and entanglement entropy in the
          absence and the presence of noise when the ratio of the
          amplitude and the frequency is fixed to be an odd integer
          (we set $A/\omega= 3$). Plots (a) and (b) show decaying
          behavior of return probability and unbounded growth of
          entanglement entropy respectively in the absence of noise. A
          suitable noise can help engineer `noise-induced' dynamical
          localization as shown by the oscillatory behavior of return
          probability and entanglement entropy respectively in (c) and
          (d). The other parameters are $L=200, \Delta=4.0$. The data
          are averaged over $100$ realizations of disorder.}
	\label{fig:fig3}
\end{figure}  

The above expression resembles Eq.~\eqref{eqn:eq25a} with the static
field $\epsilon$ replaced by an effective field $\mu\delta p$. Thus we
expect phenomena similar to dynamical localization and coherent
destruction of Wannier-Stark localization to be induced by the noisy
field. When the ratio of the amplitude to the frequency is tuned to be
an odd integer ($A/\omega=\left(2n+1\right)$), the noisy field induces
dynamical localization. Fig.\eqref{fig:fig3} shows the probability and
the entanglement entropy, for this scenario both in the absence and
presence of noise. The clean limit, as seen from Fig.\eqref{fig:fig3}
(a,b), results in delocalization behavior. On the other hand, we see
that a carefully tuned noise induces dynamical localization which is
signalled by the probability and the entropy showing oscillatory
behavior (Fig.\eqref{fig:fig3}(c,d)). This signifies the emergence of
a new kind of dynamical localization that is induced by a noisy
field. As the strength of the noise is increased, the system loses
coherence resulting in a transition to delocalization. It is worth
pointing out that in Figs.~\ref{fig:fig3}(c,d), a tendency for the
oscillations to decay is also visible, although these effects may
become important only when very long timescales are involved.

\begin{figure}[t]
	\includegraphics[scale=0.25]{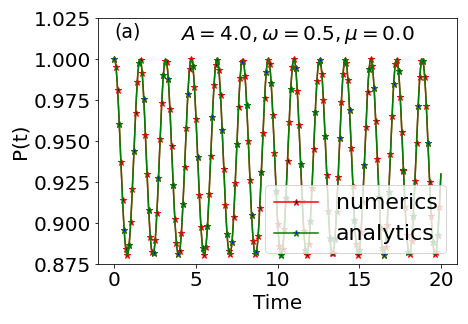}
	\includegraphics[scale=0.25]{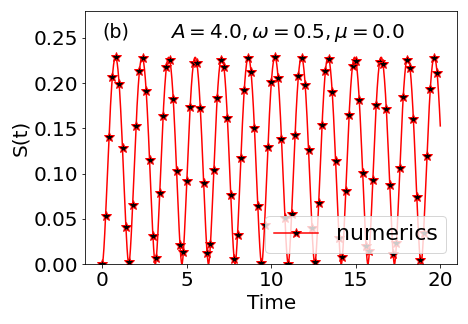}
	\includegraphics[scale=0.25]{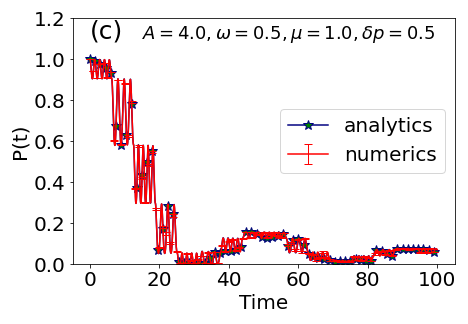}
	\includegraphics[scale=0.25]{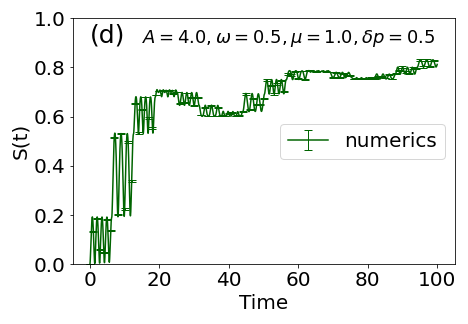}
	\caption{Plots (a) and (b) show oscillatory behavior of return
          probability and entanglement entropy corresponding to
          dynamical localization for a clean system with the tuning
          $\left(A/\omega=2n\right)$. When a suitably tuned noise is
          included delocalization behaviour can be induced as
          signalled by the decaying nature of the probability and
          unbounded growth of entanglement entropy in (c) and (d)
          respectively. We have averaged over $145$ realizations of
          disorder for a system of size $L=200$. The hopping strength
          is set to $\Delta=2.0$.}
	\label{fig:fig4}
\end{figure}

Contrastingly, the noisy field can also lead to a complete destruction
of dynamical localization when the ratio of the amplitude to the
frequency is tuned to be an even integer
($A/\omega=\left(2n\right)$). As shown in Fig.~\eqref{fig:fig4} in the
absence of the noise, the parameters of the drive lead to dynamical
localization where the probability and the entropy oscillate in time
(Fig.~\ref{fig:fig4}(a,b)). However, in the presence of an
appropriately tuned noisy field, dynamical localization is destroyed
and delocalization behavior is observed as shown in
Fig.~\eqref{fig:fig4}(c,d) where the probability and the entropy
exhibit decaying behavior and unbounded growth respectively.  These
interesting results are reminiscent of the case of a periodically
driven system together with a static field in the clean limit as
discussed in the previous section. However in this scenario, the
effects are induced by the noisy field which on average works as a
static field.
        
\section{Summary and Conclusion}\label{Summary and Conclusion}
To summarize, we study the dynamics under a noisy electric field and
examine how the phenomenon of dynamical localization in the clean
limit gets affected by the noisy field. We obtain an exact
expression for the probability propagator for a generalized field
including a static part, a time-periodic part, and a noisy part
modeled by telegraph noise. In the clean limit, we discuss the
phenomena of dynamical localization, coherent destruction of
Wannier-Stark localization, and super-Bloch oscillations with the help
of the obtained probability propagator.

In the presence of noise, we show that the dynamical localization
survives for small noise strength for some time and damps out in the
long time limit, while a larger value of the noisy field brings
decoherence to the system causing the delocalization of the
particle. When the two levels of the noise are not equi-probable, we
observe two interesting effects. In one case, with a proper tuning of
the ratio of the amplitude and the driving frequency, dynamical
localization can be destroyed incoherently, while with a different
tuning of the ratio of the amplitude and the driving frequency, we see
the emergence of dynamical localization induced by the noisy
field. Thus with a suitable tuning of the noise parameters, we are
able to go from a dynamically localized phase to a delocalized one and
vice-versa.

It is known that the clean limit of an interacting driven model can
exhibit exciting phenomena such as drive induced many-body
localization~\cite{PhysRevB.96.020201,PhysRevB.102.024201} and
coherent destruction of Stark-many-body
localization~\cite{PhysRevB.102.024201,PhysRevB.102.085133}. Thus, it
would be interesting to investigate the interplay of many-body
interactions, drive and noise to check if further noise-induced
effects can be engineered.  Another possibility for exploration would
be to consider other forms of noise, which have been studied
recently~\cite{PhysRevB.101.184308}.

\section*{Acknowledgments}
We are grateful to the High Performance Computing(HPC) facility at
IISER Bhopal, where large-scale calculations in this project were
run. We thank Sushanta Dattagupta for helpful discussions. V.T is
grateful to DST-INSPIRE for her PhD fellowship. A.S acknowledges
financial support from SERB via the grant (File Number:
CRG/2019/003447), and from DST via the DST-INSPIRE Faculty Award
[DST/INSPIRE/04/2014/002461].

\bibliography{refnew}

\onecolumngrid
\newpage
\appendix

\section{}
\subsection*{\label{app:subsec}CALCULATION OF PROBABILITY PROPAGATOR}

\vspace{2ex}

Beginning with the expression for the probability propagator (Eq.\eqref{eqn:eq15a}):
\begin{eqnarray}
 \mathcal{P}_{n}\left(t\right)&=& \left(\frac{1}{2\pi}\right)^{2}\int_{-\pi}^{\pi} dk \int_{-\pi}^{\pi} dk^{\prime} e^{-i\left(k-k^{\prime}\right)n}\times
e^{-i\int_{0}^{t}dt^{\prime}\left[\tilde{V}_{+k}\left(t^{\prime}\right)-\tilde{V}_{+k^{\prime}}\left(t^{\prime}\right)\right]},\nonumber
\end{eqnarray}
we will show how to obtain Eq.\eqref{eqn:eq10}. To
caclulate the integral that appears in the exponential of the integrand of
Eq.\eqref{eqn:eq15a}, it is helpful to recall Eq.\eqref{eqn:eq12a}:
\begin{eqnarray}
\label{eqn:appa}
e^{-i\int_{0}^{t}dt^{\prime}[V_{k}^{+}(t^{\prime})-V_{k^{\prime}}^{+}(t^{\prime})]}&=& e^{\frac{-i\Delta}{4}\int_{0}^{t}dt^{\prime}\left[ \left\lbrace e^{i(k+\eta(t^{\prime}))}+c.c.\right\rbrace-\left\lbrace e^{i(k^{\prime}+\eta(t^{\prime}))}+c.c.\right\rbrace\right]},
\end{eqnarray}
where
\begin{eqnarray}
\label{eqn:eqappb}
e^{i\eta\left(t^{\prime}\right)} &=& \left\{\begin{array}{cc}\frac{1}{2}e^{\left(iAt^{\prime}-iAnT\right)} \left[e^{(\nu-\gamma) t^{\prime}}\left(1+\hat{h}.\vec{\sigma}\right)+e^{-(\nu+\gamma) t^{\prime}}\left(1-\hat{h}.\vec{\sigma}\right)\right]; \quad 
\left(2n\pi \leq\omega t^{\prime}\leq \left(2n+1\right)\pi; \quad n=0,1,2..\right)\nonumber \\
\frac{1}{2}e^{-iAt^{\prime}+iA\left(n+1\right)T}\left[e^{(\nu-\gamma) t^{\prime}}\left(1+\hat{h}.\sigma\right)+e^{-(\nu+\gamma) t^{\prime}}\left(1-\hat{h}.\vec{\sigma}\right)\right]; \quad
\left(\left(2n+1\right)\pi <\omega t^{\prime}\leq 2\left(n+1\right)\pi \right). \end{array}\right.\\
\end{eqnarray}
Next, to simplify Eq.\eqref{eqn:appa}, we consider the following possibilities for a finite time $t$:
\begin{enumerate}[(i)]
\item $t=mT+\tau$,
\item $t=mT+\frac{T}{2}+\tau$, 
\end{enumerate}
where $0<\tau<\frac{T}{2}$ and $m$ is a non-zero positive
integer. Now, we present here our calculation for $t=mT+\tau$ and
later will generalize it for a general time $t$. We have:
\begin{eqnarray}
\label{eqn:appb1}
\int_{0}^{mt+\tau}dt^{\prime}e^{i\eta(t^{\prime})} = \frac{1}{2}\left[ \left\lbrace f_{1}(\nu) + f_{1}(-\nu)\right\rbrace + \left(\hat{h}.\vec{\sigma}\right)\left\lbrace f_{1}(\nu) - f_{1}(-\nu)\right\rbrace\right],
\end{eqnarray}
where
\begin{eqnarray}
\label{eqn:appb2}
f_{1}(\nu)&=& \left\lbrace\left( \frac{1-e^{-\left(\gamma-\nu-iA\right)T/2}}{\left(\gamma-\nu-iA\right)}+\frac{\left(1-e^{-\left(\gamma-\nu+iA\right)T/2}\right)e^{-\left(\gamma-\nu+iA\right)T/2}e^{iAT}}{\left(\gamma-\nu+iA\right)}\right)\left(\frac{1-e^{-\left(\gamma-\nu\right)mT}}{1-e^{-\left(\gamma-\nu\right)T}}\right)\right.\nonumber\\
&&\left.+\frac{e^{-\left(\gamma-\nu\right)mT}}{\left(\gamma-\nu-iA\right)}\left(1-e^{-\left(\gamma-\nu-iA\right)\left(t-mT\right)}\right)\right\rbrace
\end{eqnarray}
and
\begin{eqnarray}
\label{eqn:appb3}
f_{1}\left(-\nu\right)&=& \left\lbrace\left( \frac{1-e^{-\left(\gamma+\nu-iA\right)T/2}}{\left(\gamma+\nu-iA\right)}
+\frac{\left(1-e^{-\left(\gamma+\nu+iA\right)T/2}\right)e^{-\left(\gamma+\nu+iA\right)T/2}e^{iAT}}{\left(\gamma+\nu+iA\right)}\right)\left(\frac{1-e^{-\left(\gamma+\nu\right)mT}}{1-e^{-\left(\gamma+\nu\right)T}}\right)\right.\nonumber\\
&&\left.+
\frac{e^{-\left(\gamma+\nu\right)mT}}{\left(\gamma+\nu-iA\right)}\left(1-e^{-\left(\gamma+\nu-iA\right)\left(t-mT\right)}\right)\right\rbrace.
\end{eqnarray}

Using Eq.~\eqref{eqn:appb1} the exponential in the L.H.S. of Eq.\eqref{eqn:appa} can be expressed as:
\begin{align}
\label{eqn:appaprime}
-i\int_{0}^{mt+\tau}dt^{'}\left[V_{k}^{+}\left(t^{'}\right)-V_{k^{'}}^{+}\left(t^{'}\right)\right] &=\frac{i\Delta}{8}\left[\left(z-z^{\prime}\right)\left\lbrace f_{1}\left(\nu\right) + f_{1}\left(-\nu\right)\right\rbrace
+\left(z-z^{\prime}\right)\left(\hat{h}.\vec{\sigma}\right)
\left\lbrace f_{1}\left(\nu\right) - f_{1}\left(-\nu\right) \right\rbrace+c.c.\right],
\end{align}
where $z=e^{ik}, z^{\prime}=e^{ik^{\prime}}$. To generalize, we can write the above Eq.~\eqref{eqn:appaprime} as:
\begin{align}
\label{eqn:appb}
-i\int_{0}^{t}dt^{'}\left[V_{k}^{+}\left(t^{'}\right)-V_{k^{'}}^{+}\left(t^{'}\right)\right] &=\frac{i\Delta}{8}\left[\left(z-z^{\prime}\right)\left\lbrace f_{2}\left(\nu\right) + f_{2}\left(-\nu\right)\right\rbrace
+\left(z-z^{\prime}\right)\left(\hat{h}.\vec{\sigma}\right)
\left\lbrace f_{2}\left(\nu\right) - f_{2}\left(-\nu\right) \right\rbrace+c.c.\right].
\end{align}
where the general time $t$ can be accommodated with the aid of Heaviside step functions:
\begin{eqnarray}
\label{eqn:appc2}
f_{2}(\nu)&=& \left\lbrace\left( \frac{1-e^{-\left(\gamma-\nu-iA\right)T/2}}{\left(\gamma-\nu-iA\right)}+\frac{\left(1-e^{-\left(\gamma-\nu+iA\right)T/2}\right)e^{-\left(\gamma-\nu+iA\right)T/2}e^{iAT}}{\left(\gamma-\nu+iA\right)}\right)\left(\frac{1-e^{-\left(\gamma-\nu\right)mT}}{1-e^{-\left(\gamma-\nu\right)T}}\right)\right.\nonumber\\
&&\left.+\frac{e^{-\left(\gamma-\nu\right)mT}}{\left(\gamma-\nu-iA\right)}\left(1-e^{-\left(\gamma-\nu-iA\right)\tau}\right)+\frac{e^{-\left(\gamma-\nu\right)mT}e^{-\left(\gamma-\nu-iA\right)T/2}}{\left(\gamma-\nu+iA\right)}\left(1-e^{-\left(\gamma-\nu+iA\right)\left(t-mT-\frac{T}{2}\right)\mathcal{H}\left(t-mT-\frac{T}{2}\right)}\right)\right\rbrace,\nonumber\\
\end{eqnarray}
\begin{eqnarray}
\label{eqn:appc3}
f_{2}\left(-\nu\right)&=& \left\lbrace\left( \frac{1-e^{-\left(\gamma+\nu-iA\right)T/2}}{\left(\gamma+\nu-iA\right)}
+\frac{\left(1-e^{-\left(\gamma+\nu+iA\right)T/2}\right)e^{-\left(\gamma+\nu+iA\right)T/2}e^{iAT}}{\left(\gamma+\nu+iA\right)}\right)\left(\frac{1-e^{-\left(\gamma+\nu\right)mT}}{1-e^{-\left(\gamma+\nu\right)T}}\right)+\right.\nonumber\\
&&\left.
\frac{e^{-\left(\gamma+\nu\right)mT}}{\left(\gamma+\nu-iA\right)}\left(1-e^{-\left(\gamma+\nu-iA\right)\tau}\right)+\frac{e^{-\left(\gamma+\nu\right)mT}e^{-\left(\gamma+\nu-iA\right)T/2}}{\left(\gamma+\nu+iA\right)}\left(1-e^{-\left(\gamma+\nu+iA\right)\left(t-mT-\frac{T}{2}\right)\mathcal{H}\left(t-mT-\frac{T}{2}\right)}\right)\right\rbrace\nonumber.
\end{eqnarray}
Here $\tau=\left\{\begin{array}{cc} t-mT, & \left(t-mT\right)<\frac{T}{2}\\
\frac{T}{2}, &\left(t-mT\right)\geqslant\frac{T}{2}\end{array}\right.$, and the Heaviside step function is defined as:
$
\mathcal{H}\left(x\right)=\left\{\begin{array}{cc} 0,  & x < 0\\
1, & x \geqslant 0 \end{array}\right.$. The exponential of Eq. (\ref{eqn:appb}) can now be written in compact form as:
\begin{eqnarray}
\label{eqn:appc}
e^{-i\int_{0}^{t}dt^{\prime}[V_{k}^{+}(t^{\prime})-V_{k^{\prime}}^{+}(t^{\prime})]}&=&e^{ig_{0}\left(t\right)}e^{i\left(\textbf{H}.\vec{\sigma}\right)}\nonumber\\
&=&e^{ig_{0}\left(t\right)}\left(\mathcal{I}\cos\left(|\textbf{H}|\right)+i\left(\hat{\textbf{H}}.\vec{\sigma}\right)\sin\left(|\textbf{H}|\right)\right),
\end{eqnarray}
where we have used the identity
$e^{i\left(\vec{a}\vec{\sigma}\right)}=\left(\mathcal{I}\cos|\textbf{a}|
+i\left(\hat{a}.\vec{\sigma}\right)\sin\left(|\textbf{a}|\right)\right)$,
and defined
$H_{x}=g_{1}\left(t\right),
H_{y}=g_{2}\left(t\right)=i\delta pg_{1}\left(t\right),
H_{z}=g_{3}\left(t\right)=\delta
pg_{1}\left(t\right)+\beta\left(t\right)$ and
$|\textbf{H}|=\sqrt{g_{1}^{2}(t)+g_{2}^{2}(t)+g_{3}^{2}(t)}$. The functions $g_{0}(t), g_{1}(t),$ and $\beta\left(t\right)$ are given by:
\begin{eqnarray}
	\label{eqn:eqapph}
	g_{0}\left(t\right)&= &\frac{\Delta}{8}\left[\left(z-z^{\prime}\right)\left\lbrace f_{2}\left(\nu\right) + f_{2}\left(-\nu\right) \right\rbrace+c.c.\right]\\
\label{eqn:eqappi}
	g_{1}\left(t\right)&= &\frac{\Delta\gamma}{8}\left[\frac{\left(z-z^{\prime}\right)}{\nu}\left\lbrace f_{2}\left(\nu\right) - f_{2}\left(-\nu\right) \right\rbrace+c.c.\right]\\
	\label{eqn:eqappk}
	\beta\left(t\right)&= &\frac{i\Delta\mu}{8}\left[\left(z-z^{\prime}\right)\left\lbrace f_{2}\left(\nu\right) - f_{2}\left(-\nu\right) \right\rbrace-c.c.\right].
	\label{eqn:eqa}
	\end{eqnarray}
Hence, we arrive at the expression for the probability propagator
given in Eq.\eqref{eqn:eq10}:
\begin{eqnarray}
\label{eqn:appd}
\mathcal{P}_{n}\left(t\right)=
\left(\frac{1}{2\pi}\right)^{2}\int_{-\pi}^{\pi}dk\int_{-\pi}^{\pi}dk^{\prime}e^{-i\left(k-k^{\prime}\right)n}e^{ig_{0}\left(t\right)}
\times \left(\mathcal{I}\cos\left(|\textbf{H}|\right)+i\left(\hat{\textbf{H}}.\vec{\sigma}\right)\sin\left(|\textbf{H}|\right)\right).
\end{eqnarray}

\section{}
\subsection*{\label{app2:subsec}CALCULATION OF PROBABILITY PROPAGATOR FOR RAPID RELAXATION REGIME}

\vspace{2ex}
Here, we discuss the rapid relaxation regime in which $\gamma>>\mu, A$. In this limit $\Theta^{2}(t)>>\beta^{2}(t)$, hence we can approximate $|\textbf{H}|$ as:
\begin{equation}
|\textbf{H}| =\sqrt{g_{1}^{2}\left(t\right)+\beta^{2}\left(t\right)+2\delta pg_{1}\left(t\right)\beta\left(t\right)}\approx   g_{1}\left(t\right)\left(1+\frac{\beta^{2}\left(t\right)}{2g_{1}^{2}\left(t\right)}+\frac{\delta p\beta\left(t\right)}{g_{1}\left(t\right)}\right).
\end{equation}
This implies the following simplification to the intergand of Eq.\eqref{eqn:eq11}:
\begin{align}
\label{eqn:app2a1}
e^{ig_{0}\left(t\right)}\left(\cos\left(|\textbf{H}|\right)+i\frac{\sin(|\textbf{H}|)}{|\textbf{H}|}\Big[g_{1}\left(t\right)+\delta p \beta\left(t\right)\Big]\right)
&\approx \exp\Big(i(g_{0}\left(t\right)+|\textbf{H}|)\Big)\qquad\qquad\nonumber\\
&\approx \exp{\left(ig_{0}(t)+ig_{1}(t)+i\delta p\beta(t)+i\frac{\beta^{2}(t)}{2g_{1}(t)}\right)}.
\end{align}

Now, we consider two different cases of two-level telegraph noise based on the probability associated with both the levels in the following subsections.

\subsection{\label{app2a:subsec}Zero Bias Case \textbf{$(\delta p=0)$}}

In this subsection, we will consider the case when both the levels of noise are equally probable i.e. the case of zero bias $\left(\delta p=0\right)$. In this limit, $\nu=\sqrt{\gamma^{2}-\mu^{2}+2i\mu\gamma\delta p}\approx \left(\gamma-\frac{\mu^{2}}{2\gamma}\right)$. First, we consider the absence of square wave drive $(A=0)$, and approximate Eq.\eqref{eqn:eqapph}, \eqref{eqn:eqappi} and \eqref{eqn:eqappk} for $t=mT+\tau$:
\begin{eqnarray}
\label{eqn:app2a}
\lim_{\delta p\rightarrow 0, A\rightarrow 0}g_{0}\left(t\right)&\approx &\frac{\Delta}{4}\left[\left(z-z^{\prime}\right)\left\lbrace\left( \frac{1-e^{-\left(\frac{\mu^{2}}{2\gamma}\right)T/2}}{\left(\frac{\mu^{2}}{2\gamma}\right)}+\frac{\left(1-e^{-\left(\frac{\mu^{2}}{2\gamma}\right)T/2}\right)e^{-\left(\frac{\mu^{2}}{2\gamma}\right)T/2}}{\left(\frac{\mu^{2}}{2\gamma}\right)}\right)\right.\right.\nonumber\\
&&\left.\left. \left(\frac{1-e^{-\left(\frac{\mu^{2}}{2\gamma}\right)mT}}{1-e^{-\left(\frac{\mu^{2}}{2\gamma}\right)T}}\right)+
\frac{e^{-\left(\frac{\mu^{2}}{2\gamma}\right)mT}}{\left(\frac{\mu^{2}}{2\gamma}\right)}\left(1-e^{-\left(\frac{\mu^{2}}{2\gamma}\right)\left(t-mT\right)}\right)\right.\right.\nonumber\\
&&\left.\left.+ \left( \frac{1-e^{-\left(2\gamma\right)T/2}}{\left(2\gamma\right)}+\frac{\left(1-e^{-\left(2\gamma\right)T/2}\right)e^{-\left(2\gamma\right)T/2}}{\left(2\gamma\right)}\right) \left(\frac{1-e^{-\left(2\gamma\right)mT}}{1-e^{-\left(2\gamma\right)T}}\right)\right.\right.\nonumber\\
&&\left.\left.+
\frac{e^{-\left(2\gamma\right)mT}}{\left(2\gamma\right)}\left(1-e^{-\left(2\gamma\right)\left(t-mT\right)}\right)\right\rbrace+c.c.\right]\nonumber\\
&\approx &\frac{\Delta}{8\gamma}\left\lbrace 2\gamma t+\left(1-e^{-2\gamma t}\right)\right\rbrace\left(\cos(k)-\cos(k^{\prime})\right)\\
\label{eqn:app2aa}
\lim_{\delta p\rightarrow 0, A\rightarrow 0}g_{1}\left(t\right)&\approx &\frac{\Delta}{4}\left[\left(z-z^{\prime}\right)\left\lbrace\left( \frac{1-e^{-\left(\frac{\mu^{2}}{2\gamma}\right)T/2}}{\left(\frac{\mu^{2}}{2\gamma}\right)}+\frac{\left(1-e^{-\left(\frac{\mu^{2}}{2\gamma}\right)T/2}\right)e^{-\left(\frac{\mu^{2}}{2\gamma}\right)T/2}}{\left(\frac{\mu^{2}}{2\gamma}\right)}\right)\right.\right.\nonumber\\
&&\left.\left. \left(\frac{1-e^{-\left(\frac{\mu^{2}}{2\gamma}\right)mT}}{1-e^{-\left(\frac{\mu^{2}}{2\gamma}\right)T}}\right)+
\frac{e^{-\left(\frac{\mu^{2}}{2\gamma}\right)mT}}{\left(\frac{\mu^{2}}{2\gamma}\right)}\left(1-e^{-\left(\frac{\mu^{2}}{2\gamma}\right)\left(t-mT\right)}\right)\right.\right.\nonumber\\
&&\left.\left.- \left( \frac{1-e^{-\left(2\gamma\right)T/2}}{\left(2\gamma\right)}+\frac{\left(1-e^{-\left(2\gamma\right)T/2}\right)e^{-\left(2\gamma\right)T/2}}{\left(2\gamma\right)}\right) \left(\frac{1-e^{-\left(2\gamma\right)mT}}{1-e^{-\left(2\gamma\right)T}}\right)\right.\right.\nonumber\\
&&\left.\left.-
\frac{e^{-\left(2\gamma\right)mT}}{\left(2\gamma\right)}\left(1-e^{-\left(2\gamma\right)\left(t-mT\right)}\right)\right\rbrace+c.c.\right]\nonumber\\
&\approx &\frac{\Delta}{8\gamma}\left\lbrace 2\gamma t-\left(1-e^{-2\gamma t}\right)\right\rbrace\left(\cos(k)-\cos(k^{\prime})\right)\\
\label{eqn:app2b}
\lim_{\delta p\rightarrow 0,A\rightarrow 0}\beta\left(t\right)&\approx &\frac{i\mu\Delta}{8\gamma}\left[\left(z-z^{\prime}\right)\left\lbrace\left( \frac{1-e^{\left(-\frac{\mu^{2}}{2\gamma}\right)T/2}}{\left(\frac{\mu^{2}}{2\gamma}\right)}+\frac{\left(1-e^{-\left(\frac{\mu^{2}}{2\gamma}\right)T/2}\right)e^{-\left(\frac{\mu^{2}}{2\gamma}\right)T/2}}{\left(\frac{\mu^{2}}{2\gamma}\right)}\right)\right.\right.\nonumber\\
&&\left.\left. \left(\frac{1-e^{-\left(\frac{\mu^{2}}{2\gamma}\right)mT}}{1-e^{-\left(\frac{\mu^{2}}{2\gamma}\right)T}}\right)+
\frac{e^{-\left(\frac{\mu^{2}}{2\gamma}\right)mT}}{\left(\frac{\mu^{2}}{2\gamma}\right)}\left(1-e^{-\left(\frac{\mu^{2}}{2\gamma}\right)\left(t-mT\right)}\right)\right.\right.\nonumber\\
&&\left.\left.- \left( \frac{1-e^{-\left(2\gamma\right)T/2}}{\left(2\gamma\right)}+\frac{\left(1-e^{-\left(2\gamma\right)T/2}\right)e^{-\left(2\gamma\right)T/2}}{\left(2\gamma\right)}\right) \left(\frac{1-e^{-\left(2\gamma\right)mT}}{1-e^{-\left(2\gamma\right)T}}\right)\right.\right.\nonumber\\
&&\left.\left.-
\frac{e^{-\left(2\gamma\right)mT}}{\left(2\gamma\right)}\left(1-e^{-\left(2\gamma\right)\left(t-mT\right)}\right)\right\rbrace-c.c.\right]\nonumber\\
& \approx &\frac{-\mu\Delta}{8\gamma^{2}}\left\lbrace 2\gamma t-\left(1-e^{-2\gamma t}\right)\right\rbrace\left(\sin k-\sin k^{'}\right).
\end{eqnarray}
Addition of Eq.\eqref{eqn:app2a} and \eqref{eqn:app2aa} gives
\begin{eqnarray}
\label{eqn:app2a2}
\lim_{\delta p\rightarrow 0, A\rightarrow 0}\left(g_{0}(t)+g_{1}\left(t\right)\right)&\approx &\frac{\Delta}{4}t\left(\cos k-\cos k^{\prime}\right),
\end{eqnarray}
and from Eq.\eqref{eqn:app2aa} and \eqref{eqn:app2b}, we get
\begin{eqnarray}
\lim_{\delta p\rightarrow 0, A\rightarrow 0}\frac{i\beta^{2}\left(t\right)}{2g_{1}\left(t\right)}&\approx & \frac{i\mu^{2}\Delta}{16\gamma^{3}}\frac{\left(2\gamma t-\left[1-e^{-2\gamma t}\right]\right)\left(\sin k-\sin k^{'}\right)^{2}}{\left(\cos k - \cos k^{'}\right)}.
\end{eqnarray}
Hence, R.H.S. of Eq.\eqref{eqn:app2a1} can be expressed as 
\begin{equation}
\label{eqn:app2c}
 e^{ig_{0}\left(t\right)+ig_{1}\left(t\right)}e^{i\frac{\beta^{2}\left(t\right)}{2g_{1}\left(t\right)}}\approx e^{i\frac{\Delta}{4}t\left(\cos k-\cos k^{'}\right)}e^{i\frac{\mu^{2}\Delta}{16\gamma^{3}}\frac{\left(2\gamma t-\left[1-e^{-2\gamma t}\right]\right)\left(\sin k-\sin k^{'}\right)^{2}}{\left(\cos k - \cos k^{'}\right)}}.
\end{equation}
In the long time limit, Eq.~\eqref{eqn:app2c} is further approximated to (Eq.~\eqref{eqn:eq28a}):
\begin{eqnarray}
\label{eqn:app2ca}
e^{ig_{0}\left(t\right)+ig_{1}\left(t\right)}e^{i\frac{\beta^{2}\left(t\right)}{2g_{1}\left(t\right)}}& \approx &e^{i\frac{\Delta_{\textsf{eff}}}{4}t\left(\cos k-\cos k^{'}\right)},
\end{eqnarray}
where $\Delta_{\textsf{eff}}=\Delta\left(1+\frac{1}{2}\left(\frac{\mu}{\gamma}\right)^{2}\frac{\left(\sin k-\sin k^{\prime}\right)^{2}}{\left(\cos k-\cos k^{\prime}\right)^{2}}\right)$ is a renormalized hopping parameter.
Now, we consider an externally driven system $\left(A\neq 0\right)$.
In the rapid-relaxation limit, we can ignore
$e^{i\frac{\beta^{2}\left(t\right)}{2g_{1}\left(t\right)}}$ and
Eq.\eqref{eqn:app2a1} can be approximated as:
\begin{align}
\label{eqn:app2c1}
e^{ig_{0}\left(t\right)}\left(\cos\left(|\textbf{H}|\right)+i\frac{\sin(|\textbf{H}|)}{|\textbf{H}|}\Big[g_{1}\left(t\right)+\delta p \beta\left(t\right)\Big]\right)
&\approx \exp{\left(ig_{0}(t)+ig_{1}(t)\right)}.
\end{align}
The quantity whose exponent is taken in Eq.\eqref{eqn:app2c1} may be expressed as 
\begin{eqnarray}
\label{eqn:app2d}
\lim_{\delta p\rightarrow 0,A\neq 0}\left(g_{0}\left(t\right)+g_{1}\left(t\right)\right)& \approx & \frac{\Delta}{4}\left[\left(z-z^{'}\right)\left\lbrace\left(\frac{1-e^{-\left(\frac{\mu^{2}}{2\gamma}-iA\right)T/2}}{-iA}+\frac{\left(1-e^{-\left(\frac{\mu^{2}}{2\gamma}+iA\right)T/2}\right)e^{-\left(\frac{\mu^{2}}{2\gamma}-iA\right)T/2}}{iA}\right)\right.\right.\nonumber\\
&&\left.\left.\left(\frac{1-e^{-\left(\frac{\mu^{2}}{2\gamma}\right)mT}}{1-e^{-\left(\frac{\mu^{2}}{2\gamma}\right)T}}\right)
+\frac{e^{-\frac{\mu^{2}}{2\gamma}mT}}{-iA}\left(1-e^{-\left(\frac{\mu^{2}}{2\gamma}-iA\right)\left(t-mT\right)}\right)\right\rbrace+c.c.\right]\\  
\label{eqn:app2e}
& = &\frac{-\Delta}{4}\left[\frac{\left(t-\tau\right)}{A\frac{2\pi}{\omega}}\left(2\sin k-4\sin \left(k+A\pi/\omega\right)e^{-\frac{\mu^{2}}{2\gamma}T/2}+2e^{-\frac{\mu^{2}}{2\gamma}T}\sin k\right)\right.\nonumber\\
&&\left.+\frac{1}{A}\left(2e^{-\frac{\mu^{2}}{2\gamma}mT}\sin k-2e^{-\frac{\mu^{2}}{2\gamma}t}\sin \left(k+A\tau\right)\right)\right]+\nonumber\\
&&\frac{\Delta}{4}\left[\frac{\left(t-\tau\right)}{A\frac{2\pi}{\omega}}\left(2\sin k^{\prime}-4\sin \left(k^{\prime}+A\pi/\omega\right)e^{-\frac{\mu^{2}}{2\gamma}T/2}+2e^{-\frac{\mu^{2}}{2\gamma}T}\sin k^{\prime}\right)\right.\nonumber\\
&&\left.+\frac{1}{A}\left(2e^{-\frac{\mu^{2}}{2\gamma}mT}\sin k^{\prime}-2e^{-\frac{\mu^{2}}{2\gamma}t}\sin \left(k^{\prime}+A\tau\right)\right)\right].
\end{eqnarray}
When $\frac{A}{\omega}=$ even integer, $\sin\left(k+A\pi/\omega\right)=\sin\left(k\right)$ and in the rapid relaxation limit, we can set $e^{-\frac{\mu^{2}}{2\gamma}T}$ to unity.
Thus Eq.\eqref{eqn:app2e} leads to Eq.\eqref{eqn:eq29} which shows that the system will exhibit
dynamical localized behavior at $A/\omega=2n$ only for small values of
noise whereas for large values of noise and in the long time limit, the
system lies in the delocalized phase. In the zero noise limit,
  this expression becomes equal to our result for the square wave driven
  system (Eq.\eqref{eqn:eq24}).\\

\subsection{\label{app2a:subsec}Non-zero Bias Case $(\delta p \neq 0)$}

In this subsection, we consider the case when the two levels of noise are not equiprobable i.e. $\delta p \neq 0$. From the definition of $\nu$,
$\nu=\sqrt{\gamma^{2}-\mu^{2}+2i\mu\gamma\delta p}\approx \left(\gamma-\frac{\mu^{2}}{2\gamma}+i\mu\delta p\right)$. In the rapid relaxation limit, Eq.\eqref{eqn:app2a1} is approximated as
\begin{align}
\label{eqn:app2h1}
e^{ig_{0}\left(t\right)}\left(\cos\left(|\textbf{H}|\right)+i\frac{\sin(|\textbf{H}|)}{|\textbf{H}|}\Big[g_{1}\left(t\right)+\delta p \beta\left(t\right)\Big]\right)
&\approx \exp{\left(ig_{0}(t)+ig_{1}(t)+i\delta p \beta\left(t\right)\right)}.
\end{align}

The expression that appears in Eq.\eqref{eqn:app2h1} may be simplified as: 
\begin{eqnarray}
\label{eqn:app2h}
g_{0}\left(t\right)+g_{1}\left(t\right)+\delta p\beta\left(t\right)& =& \frac{\Delta}{4}\left[\left(z-z^{\prime}\right)\left\lbrace\left( \frac{1-e^{-\left(\gamma-\nu-iA\right)T/2}}{\left(\gamma-\nu-iA\right)}+\frac{\left(1-e^{-\left(\gamma-\nu+iA\right)T/2}\right)e^{-\left(\gamma-\nu+iA\right)T/2}e^{iAT}}{\left(\gamma-\nu+iA\right)}\right)\right.\right.\nonumber\\
&&\left.\left. \left(\frac{1-e^{-\left(\gamma-\nu\right)mT}}{1-e^{-\left(\gamma-\nu\right)T}}\right)+
\frac{e^{-\left(\gamma-\nu\right)mT}}{\left(\gamma-\nu-iA\right)}\left(1-e^{-\left(\gamma-\nu-iA\right)\left(t-mT\right)}\right)\right\rbrace+c.c.\right]\nonumber\\
&=& \frac{\Delta}{4}\left[\left(z-z^{\prime}\right)\left\lbrace\left( \frac{1-e^{-\left(\frac{\mu^{2}}{2\gamma}-i\mu\delta p-iA\right)T/2}}{\left(\frac{\mu^{2}}{2\gamma}-i\mu\delta p-iA\right)}+\frac{\left(1-e^{-\left(\frac{\mu^{2}}{2\gamma}-i\mu\delta p+iA\right)T/2}\right)e^{-\left(\frac{\mu^{2}}{2\gamma}-i\mu\delta p-iA\right)T/2}}{\left(\frac{\mu^{2}}{2\gamma}-i\mu\delta p+iA\right)}\right)\right.\right.\nonumber\\
&&\left.\left. \left(\frac{1-e^{-\left(\frac{\mu^{2}}{2\gamma}-i\mu\delta p\right)mT}}{1-e^{-\left(\frac{\mu^{2}}{2\gamma}-i\mu\delta p\right)T}}\right)+
\frac{e^{-\left(\frac{\mu^{2}}{2\gamma}-i\mu\delta p\right)mT}}{\left(\frac{\mu^{2}}{2\gamma}-i\mu\delta p-iA\right)}\left(1-e^{-\left(\frac{\mu^{2}}{2\gamma}-i\mu\delta p-iA\right)\left(t-mT\right)}\right)\right\rbrace +c.c.\right].
\end{eqnarray}

We can ignore $\frac{\mu^{2}}{2\gamma}$ in the denominators, and Eq.\eqref{eqn:app2h} with the help of further approximations can be simplified as:
\begin{eqnarray}
\label{eqn:app2i}
g_{0}\left(t\right)+g_{1}\left(t\right)+\delta p\beta\left(t\right)& =& 
\frac{\Delta}{4}\left\lbrace \frac{-2\left(t-\tau\right)\sin k}{\left(\mu \delta p+A\right)\frac{2\pi}{\omega}}+2\frac{\left(t-\tau\right) e^{-\frac{\mu^{2}}{2\gamma}\frac{\pi}{\omega}}}{\left(\mu \delta p+A\right)\frac{2\pi}{\omega}}\sin \left(k+\left(\mu\delta p+A\right)\frac{2\pi}{\omega}\right)-\right.\nonumber\\
&&\left.\frac{\left(t-\tau\right)}{\left(\mu \delta p-A\right)\frac{2\pi}{\omega}}\left[  2 e^{-\frac{\mu^{2}}{2\gamma}T/2}\sin \left(k+\left(\mu\delta p+A\right)\frac{T}{2}\right)-2 e^{-\frac{\mu^{2}}{2\gamma}T}\sin \left(k+\mu \delta p T\right) \right]\right.\nonumber\\
&&\left.-\frac{1}{\mu \delta p+A}\left( 2 e^{-\frac{\mu^{2}}{2\gamma}mT}\sin \left(k+\mu \delta p mT\right)- 
2 e^{-\frac{\mu^{2}}{2\gamma}t}\sin\left(k+\mu\delta p t+A\left(t-mT\right)\right)\right)\right\rbrace-\nonumber\\
&&\frac{\Delta}{4}\left\lbrace \frac{-2mT \sin k^{\prime}}{\left(\mu \delta p+A\right)\frac{2\pi}{\omega}}+2\frac{m Te^{-\frac{\mu^{2}}{2\gamma}T/2}}{\left(\mu \delta p+A\right)\frac{2\pi}{\omega}}\sin \left(k^{\prime}+\left(\mu\delta p+A\right)\frac{T}{2}\right)-\frac{mT}{\left(\mu \delta p-A\right)\frac{2\pi}{\omega}}\right.\nonumber\\
&&\left.\left[  2 e^{-\frac{\mu^{2}}{2\gamma}T/2}\sin \left(k^{\prime}+\left(\mu\delta p+A\right)\frac{T}{2}\right)-2 e^{-\frac{\mu^{2}}{2\gamma}T}\sin \left(k^{\prime}+\mu \delta p T\right) \right]-\frac{1}{\mu \delta p+A}\right.\nonumber\\
&&\left.\left( 2 e^{-\frac{\mu^{2}}{2\gamma}mT}\sin \left(k^{\prime}+\mu \delta p mT\right)- 
2 e^{-\frac{\mu^{2}}{2\gamma}t}\sin\left(k^{\prime}+\mu\delta p t+A\left(t-mT\right)\right)\right)\right\rbrace.
\end{eqnarray}
For small values of noise $\mu$, if the square wave drive is tuned at
the dynamical localization point $A/\omega=2n,$ Eq.\eqref{eqn:app2i} takes
the form of a periodic function (Eq.\eqref{eqn:eq30}) which signifies
dynamical localization.

If we tune the square wave drive such that $A/\omega=(2n+1)$, it
results in the following simplification of Eq.\eqref{eqn:app2i}:
\begin{eqnarray}
\label{eqn:app2j}
g_{0}\left(t\right)+g_{1}\left(t\right)+\delta p\beta\left(t\right)& =& 
\frac{\Delta}{4}\left\lbrace \frac{-2\left(t-\tau\right)\sin k}{\left(\mu \delta p+A\right)\frac{2\pi}{\omega}}+2\frac{\left(t-\tau\right) e^{-\frac{\mu^{2}}{2\gamma}\frac{\pi}{\omega}}}{\left(\mu \delta p+A\right)\frac{2\pi}{\omega}}\sin \left(k+\left(\mu\delta p\right)\frac{2\pi}{\omega}\right)-\right.\nonumber\\
&&\left.\frac{\left(t-\tau\right)}{\left(\mu \delta p-A\right)\frac{2\pi}{\omega}}\left[-2 e^{-\frac{\mu^{2}}{2\gamma}T/2}\sin \left(k+\left(\mu\delta p\right)\frac{T}{2}\right)-2 e^{-\frac{\mu^{2}}{2\gamma}T}\sin \left(k+\mu \delta p T\right) \right]\right.\nonumber\\
&&\left.-\frac{1}{\mu \delta p+A}\left( 2 e^{-\frac{\mu^{2}}{2\gamma}mT}\sin \left(k+\mu \delta p mT\right)- 
2 e^{-\frac{\mu^{2}}{2\gamma}t}\sin\left(k+\mu\delta p t+A\left(t-mT\right)\right)\right)\right\rbrace-\nonumber\\
&&\frac{\Delta}{4}\left\lbrace \frac{-2(t-\tau) \sin k^{\prime}}{\left(\mu \delta p+A\right)\frac{2\pi}{\omega}}+2\frac{(t-\tau)e^{-\frac{\mu^{2}}{2\gamma}T/2}}{\left(\mu \delta p+A\right)\frac{2\pi}{\omega}}\sin \left(k^{\prime}+\left(\mu\delta p\right)\frac{T}{2}\right)-\frac{(t-\tau)}{\left(\mu \delta p-A\right)\frac{2\pi}{\omega}}\right.\nonumber\\
&&\left.\left[  -2 e^{-\frac{\mu^{2}}{2\gamma}T/2}\sin \left(k^{\prime}+\left(\mu\delta p\right)\frac{T}{2}\right)-2 e^{-\frac{\mu^{2}}{2\gamma}T}\sin \left(k^{\prime}+\mu \delta p T\right) \right]-\right.\nonumber\\
&&\left.\frac{1}{\mu \delta p+A}\left( 2 e^{-\frac{\mu^{2}}{2\gamma}mT}\sin \left(k^{\prime}+\mu \delta p mT\right)- 
2 e^{-\frac{\mu^{2}}{2\gamma}t}\sin\left(k^{\prime}+\mu\delta p t+A\left(t-mT\right)\right)\right)\right\rbrace.
\end{eqnarray}
For small values of noise $\mu$, Eq.\eqref{eqn:app2j}
will have the same form as Eq.\eqref{eqn:eq30}.

\section{}

\subsection*{\label{app1:EE}Entanglement Entropy}

To quantify the
amount of correlations in the system, a commonly used quantifier is
the entanglement entropy which can be calculated as follows. Let
$\rho$ be the density matrix of the full system consisting of two
subsystems $A$ and $B$; the von-Neumann entropy of the subsystem $A$
is given by
\begin{equation}
\label{eqn:eq12}
S_{A}=-\text{Tr}_{B}(\rho \log_{2} \rho ).
\end{equation}
When the overall state density matrix $\rho$ is pure, $S_{A}$ is also the entanglement entropy between $A$ and $B$.

In general, the calculation of the entanglement entropy in a many-body
setting is restricted by the system size as the Hilbert space
dimension grows exponentially. However, for non-interacting systems,
this can be bypassed using a clever approach that involves only the
diagonalization of a (much smaller) correlation matrix
\cite{Peschel_2003,PhysRevB.97.125116}, thereby allowing the
exploration of large system sizes. The correlation matrix is defined
as
\begin{eqnarray}
\label{eqn:eq13}
C_{mn}= \langle c_{m}^{\dagger}c_{n}\rangle= \sum_{\alpha\epsilon A,B} \phi_{\alpha}\left(m\right)\phi_{\alpha}\left(n\right)n_{\alpha},
\end{eqnarray}
where the $\phi_{\alpha}\left(m\right)$ are the single particle
eigenstates of the Hamiltonian and $n_{\alpha}$ the corresponding
occupation numbers. The von Neumann entropy is then calculated for the
subsystem A, from the eigenvalues $\zeta_{m}$ of the subsystem
correlation matrix as
\begin{equation}
\label{eqn:eq15}
S_{A}=- \sum_{m} \left[\zeta_{m} \log \zeta_{m} + \left(1-\zeta_{m} \right) \log \left(1-\zeta_{m}\right)\right].
\end{equation}
The above result holds for the dynamics of entanglement entropy even where the eigenvalues of the sub-system correlation matrix become time-dependent.

\end{document}